\documentclass[aps,prl,twocolumn,superscriptaddress]{revtex4-2}
\usepackage{amsmath}
\usepackage{mathtools} 
\usepackage{amssymb}
\usepackage{amsthm}
\usepackage{bm} 
\usepackage{upgreek} 
\usepackage{xcolor}
\usepackage{graphicx}
\usepackage{epstopdf}
\usepackage{ulem}
\usepackage[colorlinks,linkcolor=blue,anchorcolor=blue,citecolor=blue,urlcolor=blue]{hyperref}
\usepackage{braket}
\usepackage{orcidlink}
\usepackage{siunitx}
\usepackage[english]{babel}
\usepackage{svg}

\newcommand{\nn}{{\nonumber}}

\begin{document}
\title{Strain-Engineered Electronic Structure and Superconductivity in La$_3$Ni$_2$O$_{7}$ Thin Films }

\author{Yu-Han Cao\orcidlink{0000-0002-2921-6302}} \thanks{These authors contributed equally to this work.}
\affiliation{National Laboratory of Solid State Microstructures $\&$ School of Physics, Nanjing University, Nanjing 210093, China}

\author{Kai-Yue Jiang\orcidlink{0009-0007-0395-9662}} \thanks{These authors contributed equally to this work.}
\affiliation{School of Physics and Physical Engineering, Qufu Normal University, Qufu 273165, China}

\author{Hong-Yan Lu\orcidlink{0000-0003-4715-7489}} \email[Contact author: ]{hylu@qfnu.edu.cn}
\affiliation{School of Physics and Physical Engineering, Qufu Normal University, Qufu 273165, China}

\author{Da Wang\orcidlink{0000-0003-1214-6237}} \email[Contact author: ]{dawang@nju.edu.cn}
\affiliation{National Laboratory of Solid State Microstructures $\&$ School of Physics, Nanjing University, Nanjing 210093, China}
\affiliation{Collaborative Innovation Center of Advanced Microstructures, Nanjing University, Nanjing 210093, China}

\author{Qiang-Hua Wang\orcidlink{0000-0003-2329-0306}} \email[Contact author: ]{qhwang@nju.edu.cn}
\affiliation{National Laboratory of Solid State Microstructures $\&$ School of Physics, Nanjing University, Nanjing 210093, China}
\affiliation{Collaborative Innovation Center of Advanced Microstructures, Nanjing University, Nanjing 210093, China}

\begin{abstract}
Recently, the films of the Ruddlesden-Popper (RP) nickelate superconductors, in which the (La,Pr)$_3$Ni$_2$O$_7$ system exhibits a remarkable transition temperature $T_c$ exceeding 40 K, were synthesized at ambient pressure.
We systematically investigate the band structures and electronic correlation effect to identify the key factors controlling superconductivity and pathways to enhance $T_c$.
Based on density functional theory (DFT) calculations, we construct a bilayer two-orbital ($3d_{3z^2-r^2}$ and $3d_{x^2-y^2}$) tight-binding model for a series of in-plane compression mimicking the substrate effect. We find the band energy at the $M$ point drops with the compression, leading to increase of the density of states at the Fermi level, in stark contrast to the behavior of the bulk under pressure. We then apply functional renormalization group (FRG) method to study the electronic correlation effect on the superconductivity. We find the $s_\pm$-wave pairing symmetry remains robust in the films, the same as the bulk. But somewhat surprisingly, for the films, we find $T_c$ can be enhanced by reducing the in-plane lattice constant, increasing the out-of-plane lattice constant, or further electron-doping. These findings are consistent with the itinerant picture of the superconductivity induced by spin-fluctuations and provide theoretical support for further boosting $T_c$ in future experiments.
\end{abstract}
\maketitle

{\it Introduction}. The discovery of the Ruddlesden-Popper (RP) structural bulk La$_3$Ni$_2$O$_7$  with $T_\text{c}$ reaching the liquid nitrogen temperature regime \cite{2023Nature} under high pressure, has attracted great interest recently. A large number of experimental \cite{exp-6,exp-7,exp-8,exp-9,exp-10,exp-11,exp-13,exp-16,exp-28,exp-37,exp-38,exp-41,exp-43,exp-44,exp-48,exp-49,exp-50,exp-51,exp-53,exp-54,exp-56,exp-57,exp-61,exp-68,exp-69,327pressure,exp-70,exp-71,exp-72,exp-73,exp-74,exp-75,exp-76,exp-77,exp-78,exp-79,exp-80,exp-81,exp-82,exp-83,total} and theoretical \cite{t-1,t-2,t-3,t-5,t-6,t-7,t-8,t-9,t-60,t-10,t-11,t-12,t-13,t-14,t-15,t-16,t-17,t-18,t-19,t-21,t-22,t-4,t-20,t-23,t-28,t-29,t-30,t-36,t-47,t-59,t-24,t-25,t-26,t-27,t-58,t-39,t-42,t-45,t-46,t-55,t-58,t-60,t-61,t-62,t-63,t-64,t-65,t-66,t-67,t-70,t-72,t-73,t-74,t-75,t-76,t-77,t-78,t-79,t-80,t-81,t-82,liuyq_2025_VMC327,total} studies have been conducted to investigate the superconductivity, density wave and normal state electronic properties. However, the high-pressure condition has hindered further investigations or potential applications.
Recently, two independent research teams achieved a significant breakthrough by successfully synthesizing RP bilayer thin films of La$_3$Ni$_2$O$_7$ (LNO) \cite{fe-2} and La$_{2.85}$Pr$_{0.15}$Ni$_2$O$_7$ (LPNO) \cite{fe-1}, respectively, on SrLaAlO$_4$ (SLAO) substrate, both with $T_c$ exceeding the McMillan limit at ambient pressure. This discovery has sparked a second tide in the community to investigate the RP nicklate films at ambient pressure.

By comparing these two experiments of LPNO and LNO films, we obtain the following information.
(1) The in-plane lattice constant $a=3.75$~\r{A} of the LPNO film is slightly smaller than the LNO film with $a=3.77$~\r{A}, which was suspected to be responsible for the higher $T_c$ of LPNO film. But this tendency seems to be in contrast to the bulk LNO, for which higher pressure (hence smaller $a$) is found to reduce $T_c$ \cite{327pressure}. On the other hand, the out-of-plane lattice constant $c$ of the LPNO film is longer than the LNO film. However, at first glance, weaker out-of-plane hybridization (stemed from longer $c$) seems to hamper superconductivity in many theories based on bilayer coupling. How to reconcile this paradox is actually one target of the present work.
(2) Both films exhibit tetragonal space group, consistent with the appearance of superconductivity in bulk LNO under pressure, indicating the tetragonal lattice structure is indeed crucial for superconductivity in both bulk and films \cite{fe-7}.

Further information can be obtained from later experiments.
(3) The superconductivity was found to occur within the interfacial region about 1 unit cell (UC) from the SLAO substrate \cite{fe-3}, indicating the effect of Sr (hole) doping in these two films. This hole-doping effect can explain the fact that the angle-resolved phetoemission spectroscopy (ARPES) measurement \cite{fe-6} does not observe the small $\Gamma$-centered electron-like Fermi pocket as suggested by density functional theory calculations. This may suggest that hole-doping promotes the superconductivity. But later, the synthesis of La$_2$PrNi$_2$O$_7$ ({\it not} LPNO) film achieves higher $T_c$ by removing the Sr-doping effect \cite{fe-8}, partly consistent with the disappearance of the $M$-centered hole-like Fermi pocket \cite{fe-9}. This otherwise may suggest that electron-doping promotes superconductivity. Regarding to these progresses, the doping effect on electronic structure and superconductivity desires more systematic studies.
(4) The ARPES study has revealed a nodeless superconducting gap on the Fermi pockets near the $\Gamma-M$ direction as dominated by the $d_{x^2-y^2}$ orbital \cite{fe-4}, suggesting $s$- or $s_\pm$-wave pairing. Although this pairing symmetry is consistent with the bulk as predicted by many theoretical studies, the underlying superconducting mechanism of these theories differs significantly. The pressure or substrate dependence of both bulk and film materials hence provides a unique way to distinguish the superconducting mechanisms.

\begin{figure*}
    \includegraphics[width=1\linewidth]{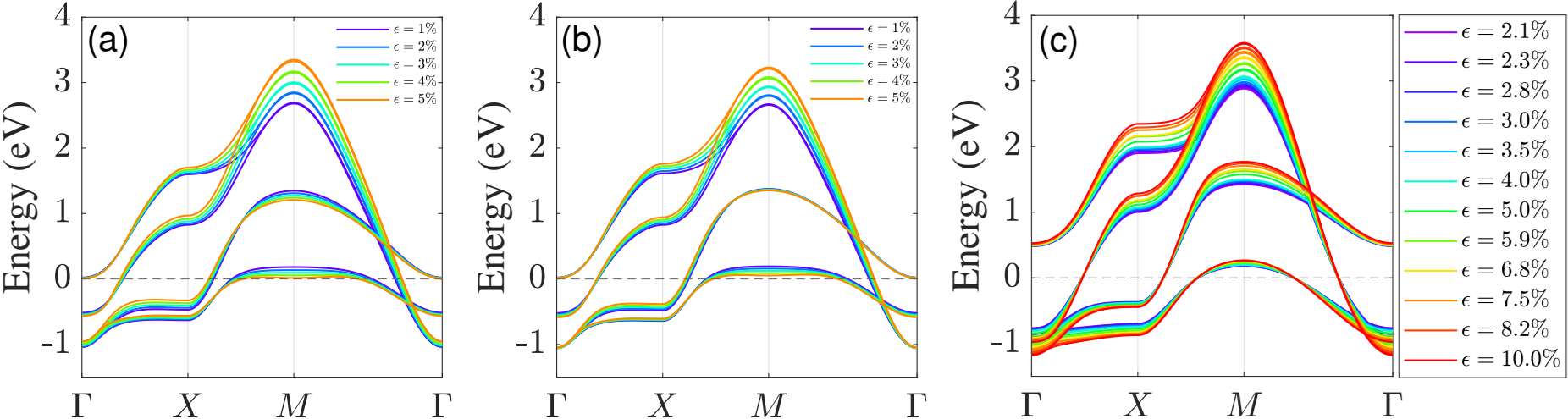}
	\caption{ (a) Tight-binding band structures with $n = 1.25$ (electron number per Ni atom) under different in-plane compression ratio $\epsilon$ corresponding to the LPNO films, by fixing the out-of-plane expansion ratio $\epsilon_c=0.5 \epsilon$. (b) is similar to (a) but for the LNO films by fixing $\epsilon_c=0$. (c) is similar to (a) but for the LNO bulk with $n = 1.5$ (converted from Ref.~\cite{LNO_pressure_PRL_2025}).}
    \label{eb}
\end{figure*}

In this work, we systematically investigate the band structures and electronic correlation induced superconductivity of the LPNO and LNO films, aiming to uncover the key factors controlling $T_c$.
Based on DFT calculations, we construct a bilayer two-orbital ($d_{3z^2-r^2}$ and $d_{x^2-y^2}$) tight-binding model. The tight-binding parameters and band structures are compared with the bulk to uncover their differences. We then apply functional renormalization group (FRG) to study the electronic correlation effect on the superconductivity. We find the $s_\pm$-wave pairing symmetry remains robust in the films, the same as its bulk counterpart. But somewhat surprisingly, for the films, we find $T_c$ can be enhanced by (1) reducing the in-plane lattice constant, (2) increasing the out-of-plane lattice constant, or (3) further electron-doping, which can be understood by the itinerant picture of the superconductivity. These findings provide theoretical support for further enhancing $T_c$ in future experiments.




{\it Electronic structure $\&$ tight-binding model}.
The electronic structures of the LPNO and LNO films are calculated by DFT as implemented in the Vienna ab initio simulation package (VASP) on a half-UC thick La$_3$Ni$_2$O$_7$ film with the $I4/mmm$ space group. Due to the small concentration of Pr in LPNO, the two films are solely distinguished by different out-of-plane lattice constants while discarding the Pr doping.
In experimental studies, LPNO and LNO thin films exhibit the approximately 2\% of in-plane compression ratio compared to LNO bulk under ambient pressure, while the out-of-plane lattice constants show 1\% elongation and remain virtually unchanged, respectively \cite{fe-1,fe-2}. Therefore, we set the in-plane compression ratio $\epsilon$ at $1\%$ increments from $1\%$ to $5\%$ for both films. Correspondingly, the out-of-plane expansion ratio $\epsilon_{c}$ are set at $0.5\%$ increments from $0.5\%$ to $2.5\%$ for LPNO, and fixed at $0\%$ for LNO, respectively.
With the method of maximally localized Wannier functions, we extract a bilayer two-orbital tight-binding model
\begin{equation}
H_0=\sum_{i \delta, a b, \sigma} t_\delta^{a b} c_{i a \sigma}^{\dagger} c_{i+\delta b \sigma}+\sum_{i a \sigma} \varepsilon_a c_{i a \sigma}^{\dagger} c_{i a \sigma},
\end{equation}
where $t_\delta^{a b}$ is the hopping matrix element between the $a$ orbital on site $i$ and the $b$ orbital on site $i$ + $\delta$, $\sigma$ represents spin, and $\varepsilon_a$ is the on-site energy of the $a$-orbital. For simplicity, we use $x$ and $z$ to denote $3d_{x^2-y^2}$ and $3d_{3z^2-r^2}$ orbitals, respectively. The full DFT bands and comparisons with the Wannier bands are shown in Supplemental Material (SM) \cite{SM}.
All the tight-binding parameters are listed in Table S1 and S2 \cite{SM} for the two films, respectively.

\begin{figure}[b]
	\includegraphics[width=\linewidth]{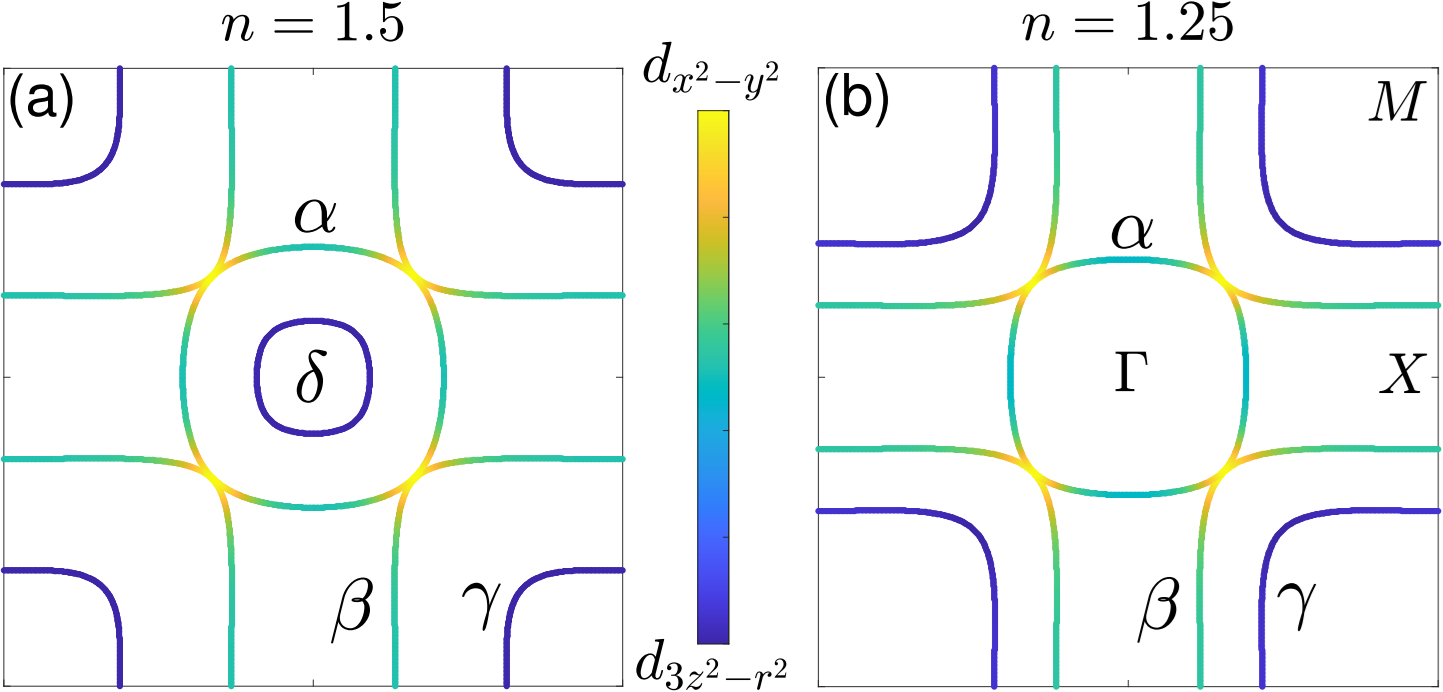}
	\caption{Fermi surfaces for the LPNO film under $\epsilon = 2\%$ for $n = 1.5$ in (a) and $n = 1.25$ in (b). The colors represent the orbital weights. The results for the LNO film are similar and not shown.}
	\label{fs}
\end{figure}

\begin{figure*}
    \includegraphics[width=1\linewidth]{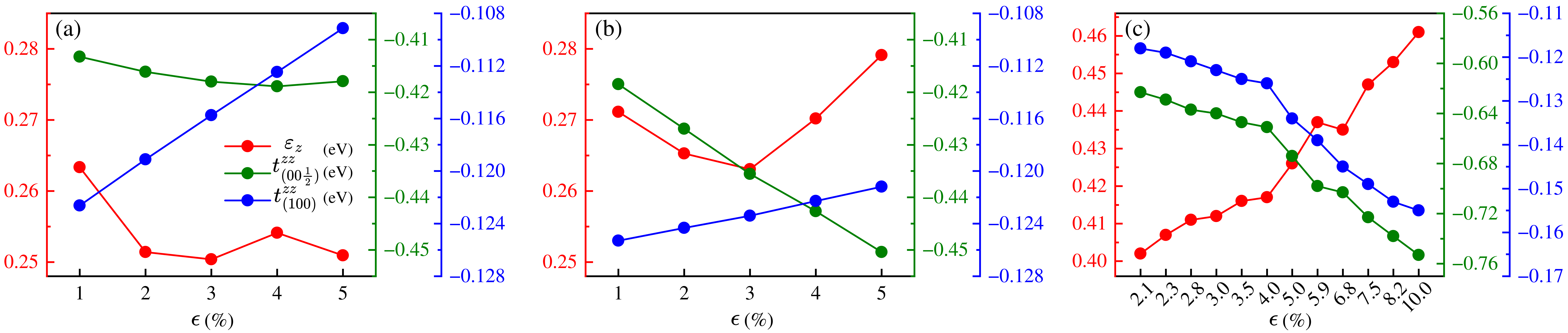}
	\caption{ The onsite energy $\varepsilon_{z}$ and hopping integrals $t^{zz}_{(00\frac{1}{2})}$ and $t^{zz}_{(100)}$ are plotted with respect to $\epsilon$ for the LPNO film (a), LNO film (b) and bulk LNO (c), respectively.}
    \label{ztb}
\end{figure*}

The tight-binding band structures of LPNO and LNO films under different compression ratio $\epsilon$ are presented in Fig.~\ref{eb}(a) and Fig.~\ref{eb}(b), respectively. The Fermi energy corresponds to $1.25$ electrons per Ni atom (labeled by $n$). As a comparison, the band structures of the bulk LNO with $n=1.5$ for different $\epsilon$ (converted from hydrostatic pressure) \cite{LNO_pressure_PRL_2025} are shown in Fig.~\ref{eb}(c). Although the total band width increases for both films and bulk with increasing $\epsilon$, the low energy (near the Fermi energy) bands exhibit quite different features. First, for both films, the energy band above the Fermi level at $\Gamma$ drops very close to the Fermi energy, leading to a small electron-like Fermi pocket upon electron-doping, as shown in Fig.~\ref{fs}(a). Second, the hole band around $M$ of both films becomes more flat than the bulk. Third, maybe the most intriguing feature of the films is the drop of the $M$-band with increasing $\epsilon$, which is opposite to the tendency in the bulk. All these features lead to increase of the density of states (DOS) at the Fermi level in the films, reflecting the distinct effect of the in-plane strain versus the hydrostatic pressure on the band structure.

Since the in-plane lattice constant $a$ is reduced for both films and bulk under pressure, the in-plane hopping amplitude between $d_{x^2-y^2}$-orbital is enhanced for all these cases (see Table S1 and S2 in SM \cite{SM}). But because the out-of-plane lattice constant $c$ increases for LPNO film, keeps invariant for LNO film, and drops for the bulk, the hopping integrals between $d_{3z^2-r^2}$-orbital differ significantly among these three cases. In Fig.~\ref{ztb}, we plot $t_{(00\frac12)}^{zz}$, $t_{(100)}^{zz}$ and $\varepsilon_z$ versus the in-plane compression ratio $\epsilon$ for the two films and the bulk, respectively. For the bulk, both $t_{(00\frac12)}^{zz}$ and $t_{(100)}^{zz}$ are enhanced with increasing $\epsilon$, hence, reducing $c$, as anticipated. But for both films, with increasing $\epsilon$ (reducing $a$), $t_{(100)}^{zz}$ is weakened unexpectedly. Moreover, for the LNO film, $t_{(00\frac12)}^{zz}$ is enhanced although $c$ keeps invariant, while for the LPNO film, $t_{(00\frac12)}^{zz}$ is slighly enhanced although $c$ increases. These counter-intuitive results come from the narrowing of the effective Wannier orbital carrying the $d_{3z^2-r^2}$ symmetry with increasing $\epsilon$, uncovering a fundamental difference between the films upon in-plane strain and the bulk under hydrostatic pressure. With these results, we can understand the drop of $M$-band with increasing $\epsilon$.
At $M$, the lowest band comes from the bonding state of the $z$-orbitals, with the energy given by $\varepsilon_{z} + t^{zz}_{(00\frac{1}{2})} - 4t^{zz}_{(100)}$ up to the nearest neighbor hopping. For the bulk under pressure, the $M$-band grows up mainly due to the increase of $\varepsilon_z$, while for the films, the $M$-band drops mainly due to the hopping changes.
The anomalous variation of $t^{zz}_\delta$ under in-plane strain is a peculiar property of the films \cite{ft-8,ft-9,ft-6,ft-12}.

{\it Electronic correlation $\&$ superconductivity}.
Next, we investigate electronic correlation induced superconductivity by considering the atomic multi-orbital Coulomb interactions
\begin{align}
H_I= & \sum_{i, a<b, \sigma \sigma^{\prime}}\left(U^{\prime} n_{i a \sigma} n_{i b \sigma^{\prime}}+J_H c_{i a \sigma}^{\dagger} c_{i b \sigma} c_{i b \sigma^{\prime}}^{\dagger} c_{i a \sigma^{\prime}}\right) \nn\\
& +\sum_{i a} U n_{i a \uparrow} n_{i a \downarrow}+\sum_{i, a \neq b} J_P c_{i a \uparrow}^{\dagger} c_{i a \downarrow}^{\dagger} c_{i b \downarrow} c_{i b \uparrow} ,
\end{align}
where $U$ is the intra-orbital Hubbard repulsion, $U'$ is the inter-orbital Coulomb interaction, $J_H$ is the Hund's coupling, and $J_P$ is the pair hopping interaction. As usual, we adopt the Kanamori relations: $U=U'+2J_H$ and $J_H=J_P$ \cite{KanamoriRelations}. The two independent interaction parameters are chosen as $U = 3$~eV and $J_H = 0.3$~eV throughout this work.

To investigate electronic correlation properties, we employ the singular-mode FRG (SM-FRG) method, which is an unbiased approach for analyzing system instabilities by studying the flow of the one-particle-irreducible four-point vertex $\Gamma_{\Lambda}$ versus a running infrared cutoff energy scale $\Lambda$ while neglecting the self-energy and frequency-dependence in our approach. The vertex $\Gamma_\Lambda$ can be reexpressed as scattering matrices between fermion bilinears in superconductivity (SC), spin-density-wave (SDW), and charge-density-wave (CDW) channels. Through iterative solution of the FRG flow equation, we monitor the negative leading singular value $S_{\Lambda}$ of the scattering matrices versus $\Lambda$. At $\Lambda_c \sim T_c$, the first divergence of the $S_{\Lambda}$ among the three channels indicates an instability to form a long range order. More technical details can be found in Refs.~\cite{Wang_PRB_2012, Wang_PRB_2013, Tang_PRB_2019, Yangqg_prb_2022, t-2, Yangqg_prb_2024_4310, Yangqg_prb_2024_CBO, LNO_pressure_PRL_2025}.

\begin{figure}[t]
	\includegraphics[width=\linewidth]{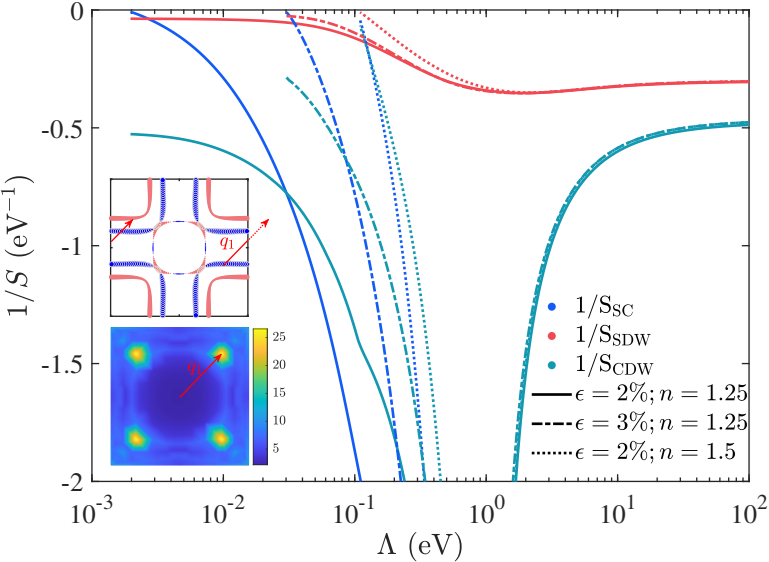}
	\caption{SM-FRG flows of $S^{-1}$ versus $\Lambda$ in the SC, SDW and CDW channels of LPNO film, respectively, at $(\epsilon,n)$ = (2\%, 1.25), (3\%, 1.25) and (2\%, 1.5). The upper subset plots the gap function on the Fermi surfaces with the pink (blue) color indicating the positive (negative) sign and the size indicating the magnitude of the gap function. The lower subset presents the leading negative $|S(\0q)|$ in the SDW channel. Both the subsets are obtained at $(\epsilon,n)=(2\%, 1.25)$.}
	\label{flow}
\end{figure}

In Fig.~\ref{flow}, we present typical SM-FRG flows of $S^{-1}$ versus $\Lambda$ for three values of $(\epsilon,n)=(2\%,1.25)$, $(3\%, 1.25)$ and $(2\%,1.5)$, respectively, for the LPNO film. The results for the LNO film are similar and can be found in Fig.~S3 in the SM \cite{SM}. At high $\Lambda$, the interactions are not renormalized, and thus the flows for the three cases of $(\epsilon,n)$ merge together. Due to the bare repulsive Coulomb interactions, the SDW channel dominates while the attractive SC channel has not been established. As $\Lambda$ reduces, the CDW channel is significantly weakened due to Coulomb screening while the SDW channel keeps almost the same. When $\Lambda$ reduces below $\sim 1$~eV, off-site spin correlations are established and enhance the SDW channel, which further induce the CDW and SC channels due to inter-channel overlap in the spirit of spin-fluctuation induced superconductivity. After the attractive SC interaction is established, the SC channel grows up faster than the other two channels due to Cooper instability, also known as the Kohn-Luttinger mechanism \cite{Kohn-Luttinger_1965}.
For $(\epsilon,n) = (2\%, 1.25)$, the SC channel diverges ($S^{-1}$ approaches zero) first, indicating that the system undergoes a phase transition into the superconductivity phase. By analyzing the eigenmode of the SC channel, we find that the real-space pairing is mainly dominated by local inter-layer pairing between $d_{3z^2-r^2}$ orbitals. More pairing components for different $(\epsilon,n)$ can be found in Fig. S4, Table S3 and Table S4 in the SM \cite{SM}.
In the upper subset of Fig.~\ref{flow}, we plot the superconducting gap function projected onto the Fermi surface, from which the $s_\pm$-wave pairing symmetry is clearly seen.
To analyze the pairing mechanism, we plot the momentum dependence $S(\0q)$ of the SDW channel in the lower inset of Fig.~\ref{flow}, which exhibits peaks at momentum $\mathbf{q_1}=(5/8,5/8)\pi$, up to $C_{4v}$ symmetry. The momentum $\mathbf{q_1}$ can connect the Fermi surface with opposite sign gap functions, showing that the superconductivity is induced by spin fluctuations. For $(\epsilon,n) = (3\%, 1.25)$, the SM-FRG flows are qualitatively similar to the above, but with a higher critical energy scale $\Lambda_c$ and stronger SDW channel. This indicates that larger $\epsilon$ enhances spin fluctuations, and thereby promotes $T_c$. For $(\epsilon,n) = (2\%, 1.5)$, although the SC fluctuation is stronger, the SDW channel diverges first and prevents the occurrence of superconductivity.

\begin{figure}
	\includegraphics[width=\linewidth]{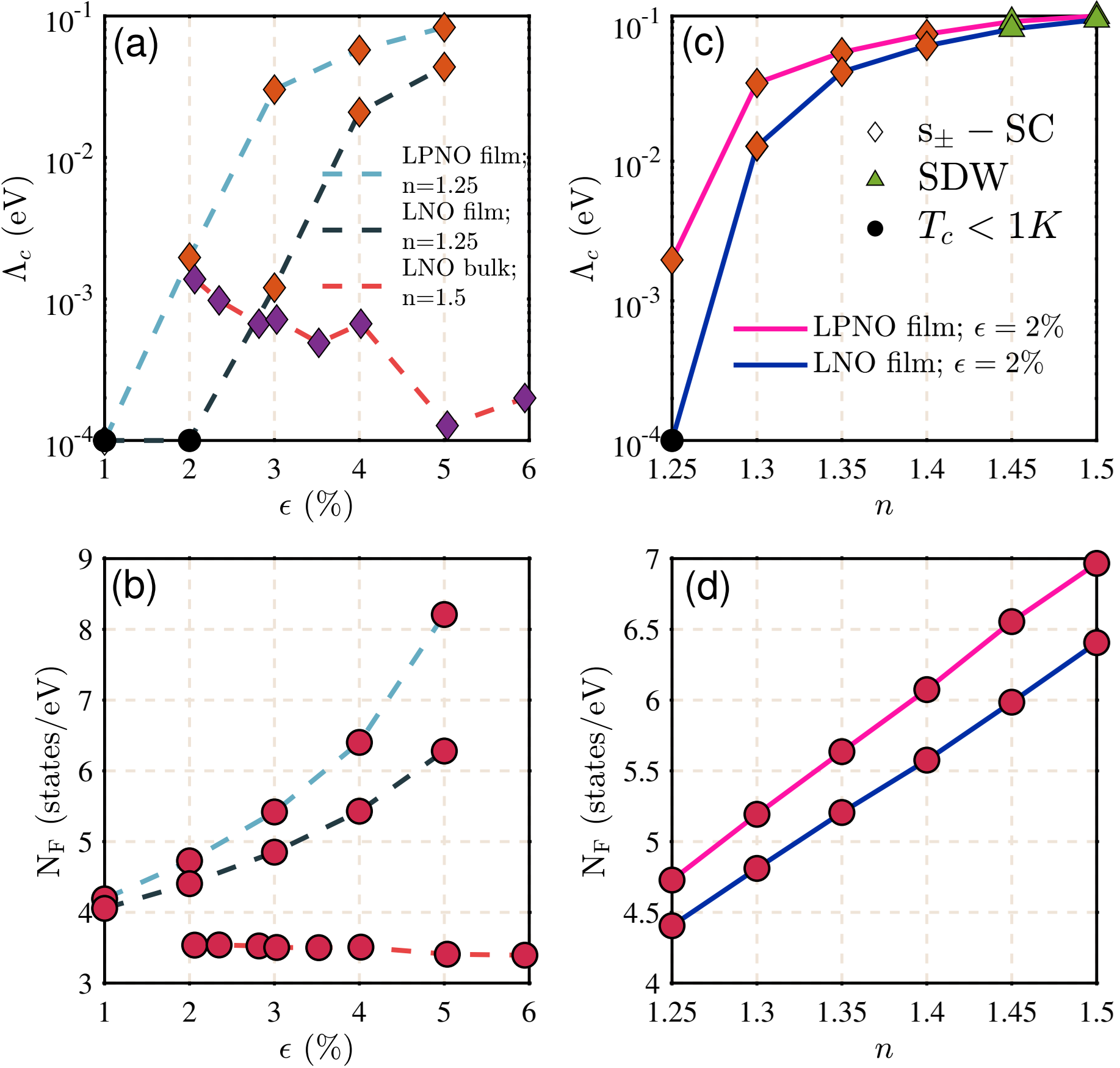}
	\caption{(a) Phase diagram of $T_c$ ($T_c \sim \Lambda_c$) versus $\epsilon$ of the LPNO film, LNO film and LNO bulk, respectively. (b) DOS at the Fermi level ($N_\mathrm{F}$) corresponding to (a). (c) Phase diagram of $T_c$ versus $n$ of the LPNO and LNO films under $\epsilon$ = 2\%. (d) DOS $N_\mathrm{F}$ corresponding to (c). }
	\label{Tc}
\end{figure}


We have performed more calculations for a series of the in-plane strain $\epsilon$ and electron density $n$. The resulting $T_c$ versus $\epsilon$ by fixing $n=1.25$ for both films are plotted in Fig.~\ref{Tc}(a). As a comparison, the FRG results for the bulk are also presented \cite{LNO_pressure_PRL_2025}, which are consistent with the experiment \cite{327pressure}.
Strikingly, we find that the $T_c$ of both films are enhanced with increasing $\epsilon$, in stark contrast to the suppression of $T_c$ in the bulk. Such an opposite trend shows a fundamental distinction between the two ways of band engineering via pressure. How to understand this difference? We attribute it mainly to the qualitatively different behavior of the DOS at the Fermi level $N_F$, as shown in Fig.~\ref{Tc}(b). As discussed above, it stems directly from the distinct energy shift of the $M$-band with $\epsilon$.
Another striking feature is the theoretical $T_c$ of the films can be higher than the bulk if higher in-plane strain can be realized experimentally on suitable substrate. Of course, the theoretical $T_c$ should be understood as the pairing temperature. Phase fluctuation in the 2D films may further reduce it to the Kosterlitz–Thouless transition temperature.

It is also interesting to compare the two films directly. Note that the LPNO film has higher $T_c$ than the LNO film for the same $\epsilon$, due to higher DOS for the LPNO film although its out-of-plane lattice constant $c$ is larger. This result is actually consistent with the experiments \cite{fe-1,fe-2}.
We also examine the doping dependence of $T_c$, which is plotted versus $n$ at fixed $\epsilon = 2\%$ in Fig.~\ref{Tc}(c). Upon electron-doping with increasing $n$, the value of $T_c$ grows up for both films. Again, this is consistent with the variation of the DOS $N_\mathrm{F}$ as shown in Fig.~\ref{Tc}(d). Of course, as shown above, for large enough $n$, the system enters the SDW phase and hampers the formation of superconductivity.
Here, both the strain and doping dependences of $T_c$ can be understood by the itinerant picture of superconductivity. If confirmed by experiments, these features can be used to identify the itinerant pairing mechanism in RP nickelate superconductors.

{\it Conclusion}.
We systematically investigate the electronic band structures and correlation effects of the LPNO and LNO films. By DFT, we construct a bilayer two-orbital tight-binding model, uncovering the distinction of films relative to their bulk counterpart. Then by FRG, we investigate the electronic correlation induced superconductivity. We find the $s_\pm$-wave pairing symmetry remains robust in the films, the same as the bulk. For the films, we find $T_c$ can be enhanced by reducing the in-plane lattice constant, increasing the out-of-plane lattice constant, or electron-doping. These findings support  the itinerant picture of the superconductivity and offer theoretical support for further increasing $T_c$ of the bilayer RP nickelates in future experiments.

Finally, it should be mentioned that our calculations are performed on the LPNO \cite{fe-1} and LNO \cite{fe-2} films, and may not be workable for the La$_2$PrNi$_2$O$_7$ film \cite{fe-8}, for which the $M$-pocket is not observed by ARPES \cite{fe-9}. If the $M$-pocket falls below the Fermi level, then we expect hole-doping can make it reappear to enhance $T_c$ following the itinerant picture of superconductivity, since the contribution of the $M$-pocket to the DOS on the Fermi level is considerably large.

{\it Acknowledgments}.
This work is supported by National Key R\&D Program of China (Grants No. 2024YFA1408100, No. 2022YFA1403201), National Natural Science Foundation of China (Grants No. 12074213, No. 12374147, No. 12274205, No. 92365203), and Major Basic Program of Natural Science Foundation of Shandong Province (Grant No. ZR2021ZD01).

{\it Data availability}.
The data that support the findings of this article are not publicly available. The data are available from the authors upon reasonable request.


\begin{widetext}
\vspace{20pt}
\begin{center}
{\LARGE Supplemental Materials}
\end{center}

\section{FIRST-PRINCIPLES CALCULATION RESULTS}

\subsection{First-principles calculation details}
The calculations of structural relaxations and band structures for half-UC LNO thin film under different compressive strains are performed by density functional theory (DFT) calculations as implemented in Vienna {\it ab initio} simulation package (VASP) \cite{VASP}. The projector augmented-wave (PAW) \cite{PAW} method and the generalized gradient approximation (GGA) of Perdew-Burke-Ernzerhof (PBE) \cite{PBE} exchange-correlation functional are adopted. The La 5$s^{2}$5$p^{6}$5$d^{1}$6$s^{2}$, Ni 3$d^{8}$4$s^{2}$, and O 2$s^{2}$2$p^{4}$ configurations are treated as valence electrons. The cutoff energy of plane-wave is set as 500 eV in all calculations. The total energies and force are well converged within 10$^{-6}$ eV and 10$^{-2}$ eV/\AA, respectively. The $\Gamma$-centered $9\times9\times1$ $k$-point grid is used for both the structural relaxation and self-consistent calculations. Subsequently, to construct a bilayer two-orbital tight-binding model, the maximally localized Wannier functions (MLWF) \cite{Wannier1,Wannier2} of LNO thin film with $3d_{3z^2-r^2}$ and $3d_{x^2-y^2}$ orbitals on each Ni atom are extracted by the Wannier90 code \cite{Wannier3}, and the $18\times18\times1$ $k$-point grid is used for wannierization. The good overlapping between DFT and Wannier band structures indicates the high quality of Wannier functions, which is displayed in Fig.~\ref{D-W}. In addition, the main tight-binding parameters extracted by Wannier90 are displayed Fig.~\ref{TB-P}, Table~\ref{T2} and Table~\ref{T1}.

\begin{figure*}[htbp]
    \centering
    \includegraphics[width=1\linewidth]{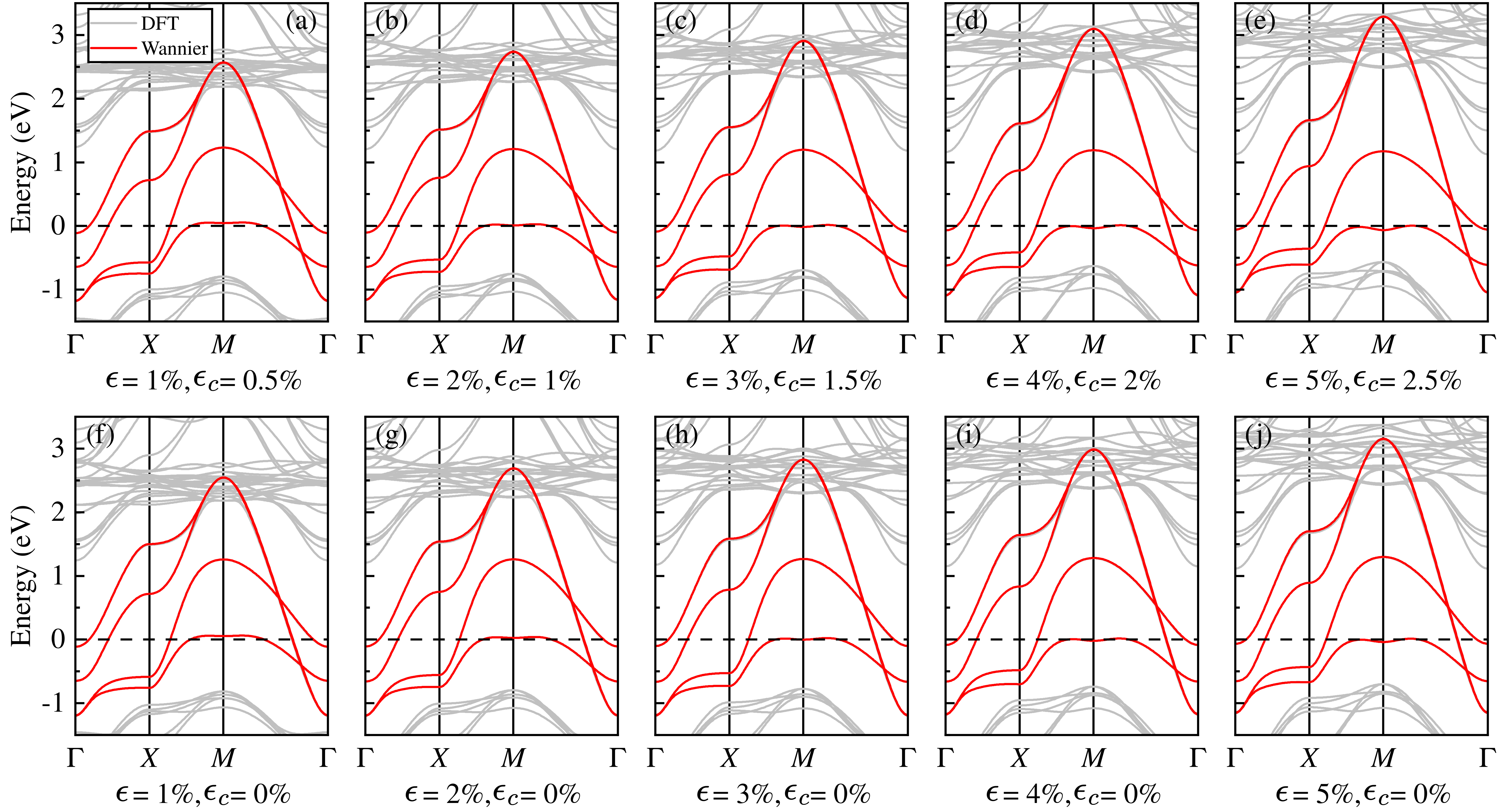}
    \caption{Comparison of DFT (grey line) and Wannier (red line) band structure corresponding to the LPNO (a-e) and LNO (f-j) films under different compressive strains.}
    \label{D-W}
\end{figure*}

\clearpage

\subsection{Main tight-binding parameters under different compressive strains}

\begin{figure*}[htbp]
    \centering
    \includegraphics[width=0.75\linewidth]{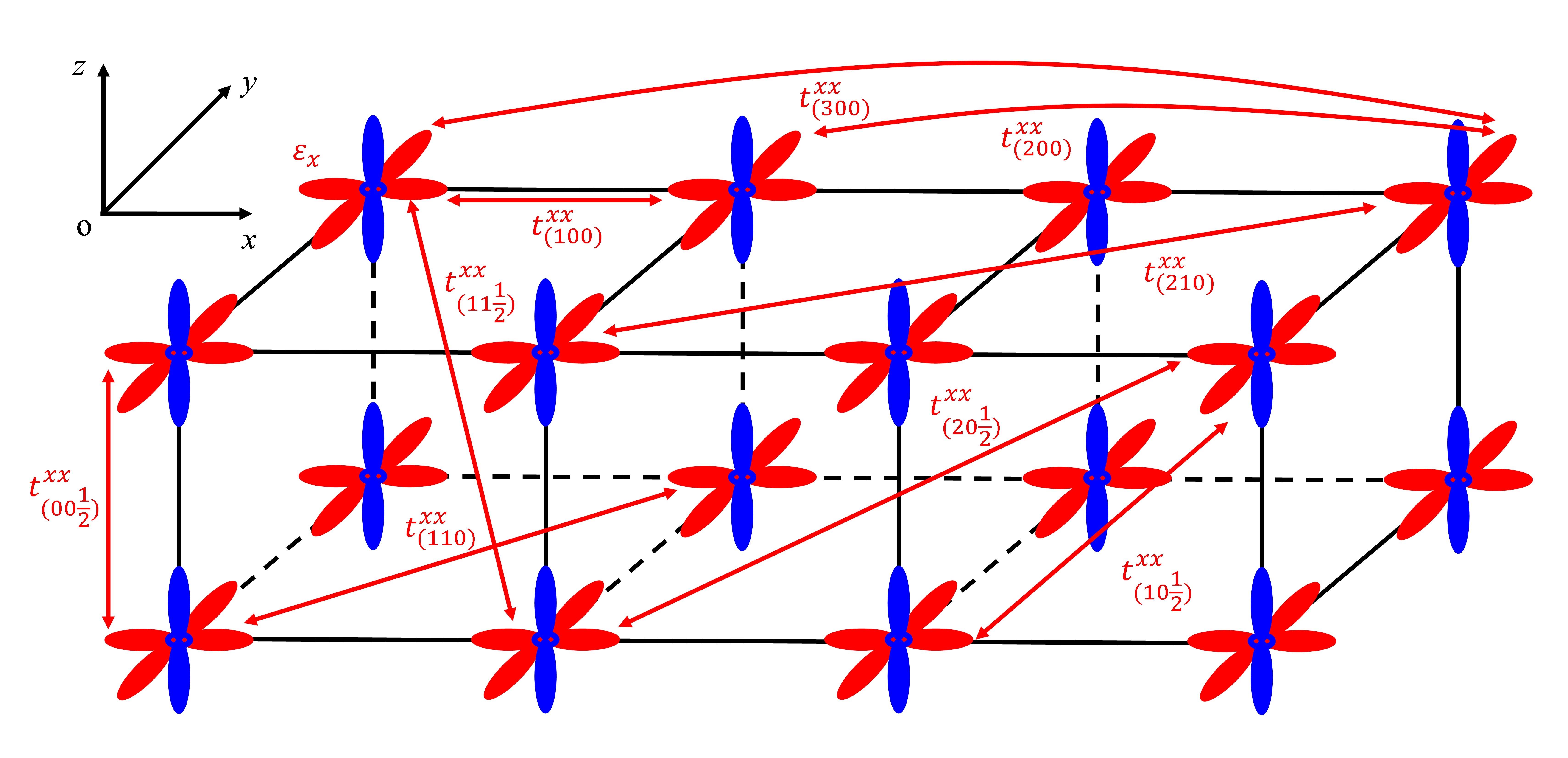}
    \includegraphics[width=0.75\linewidth]{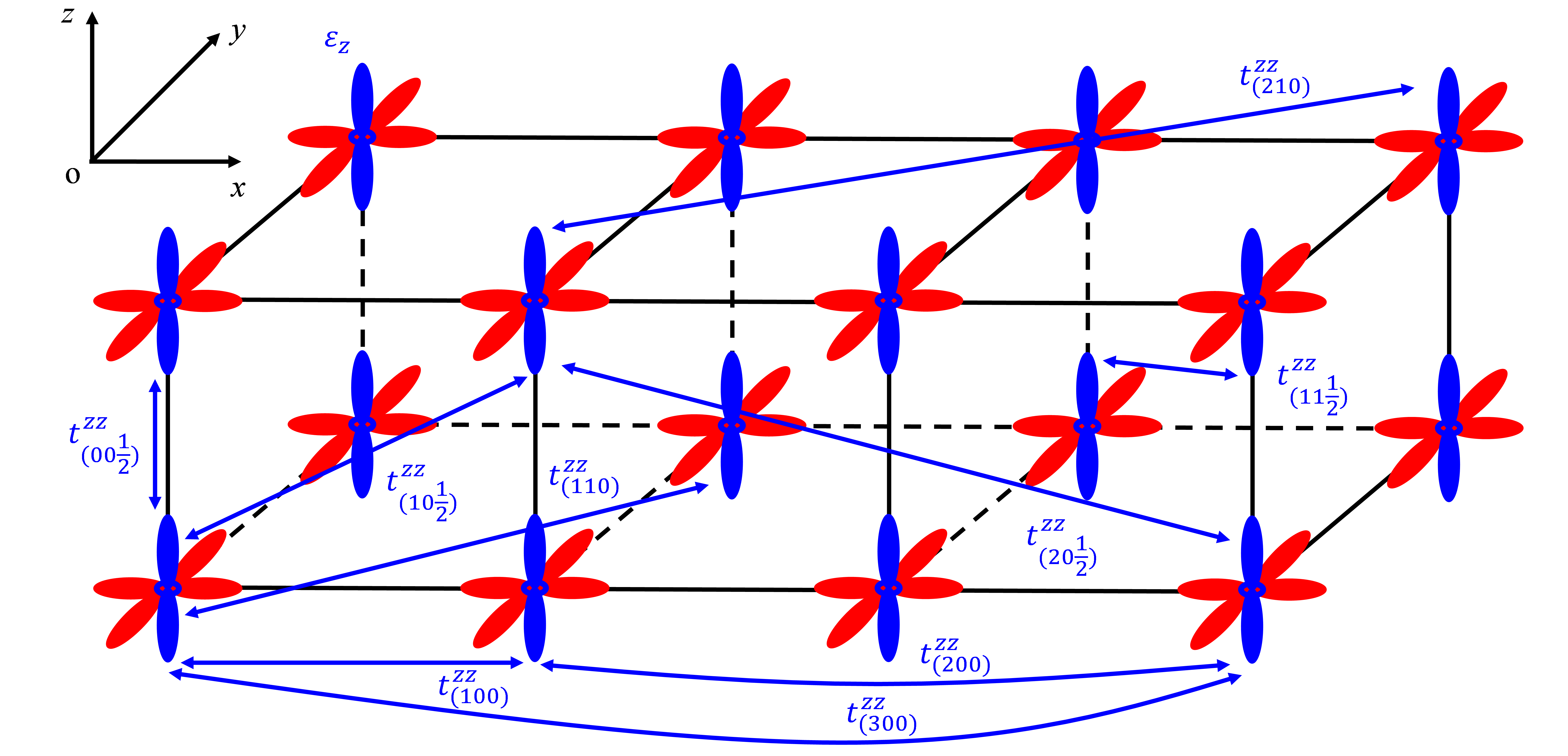}
    \includegraphics[width=0.75\linewidth]{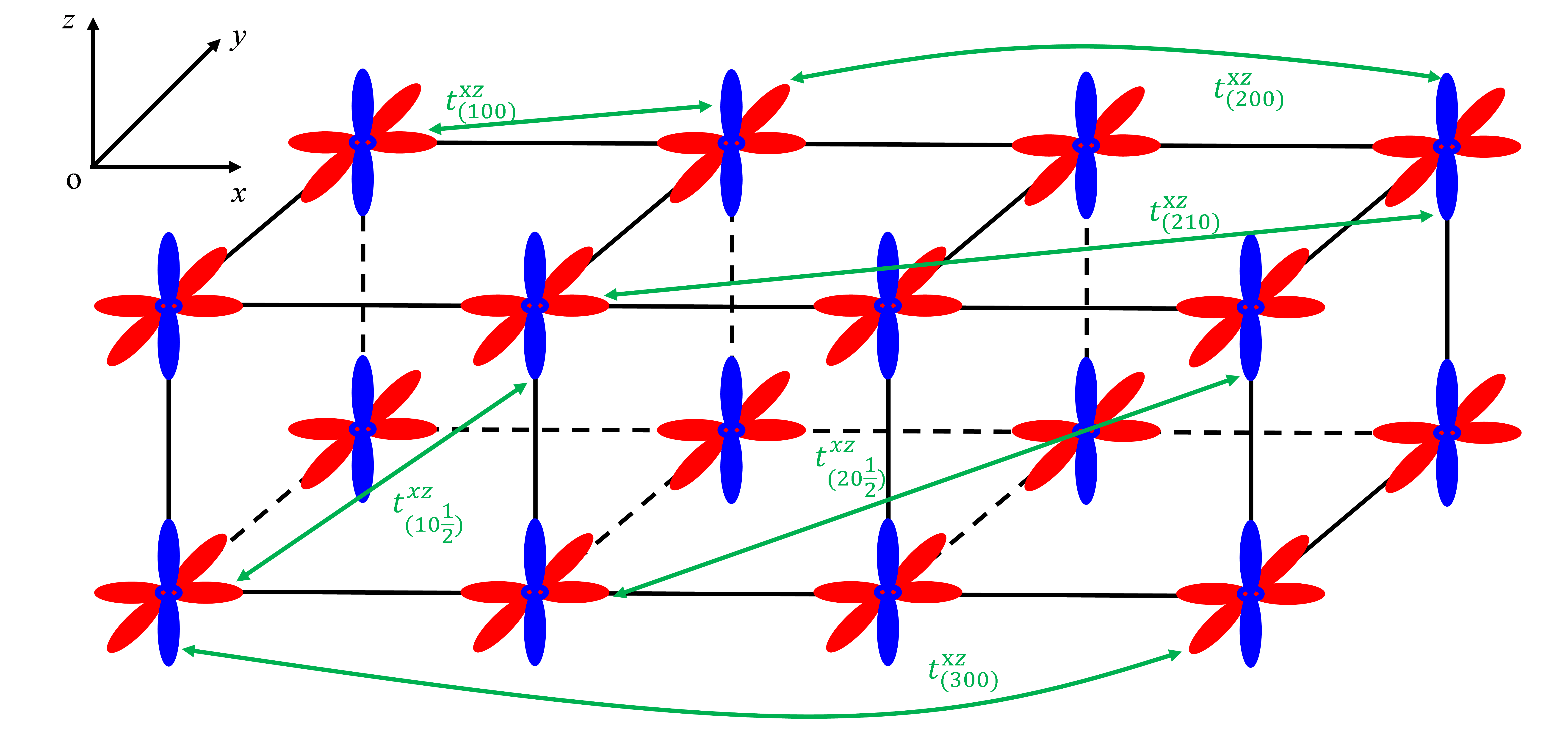}
    \caption{Schematic diagram for all parameters of the tight-binding model.}
    \label{TB-P}
\end{figure*}

\clearpage

    \begin{table}[htbp]
	\caption{On-site energies $\varepsilon_a$ and hopping integrals $t_\delta^{ab}$ of the bilayer two-orbital tight-binding model corresponding to LPNO thin film under different compressive strains. Here, $x$ and $z$ denote the $3d_{x^2-y^2}$ and $3d_{3z^2-r^2}$ orbital, respectively. Note that the vertical interlayer distance is assigned as $\frac12$. The unit of $\varepsilon_a$ and $t_\delta^{ab}$ is eV.}
	\centering
	\setlength{\tabcolsep}{10pt}
	\renewcommand{\arraystretch}{1.5}
	\begin{tabular}{cccccc}
		\hline
		\hline
		& $\epsilon$=1\%, $\epsilon_c$=0.5\% & $\epsilon$=2\%, $\epsilon_c$=1\% & $\epsilon$=3\%, $\epsilon_c$=1.5\% & $\epsilon$=4\%, $\epsilon_c$=2\% & $\epsilon$=5\%, $\epsilon_c$=2.5\% \\ \hline
		$\varepsilon_x$ & 0.612463 & 0.712498 & 0.821973 & 0.948081 & 1.078286 \\
		$\varepsilon_z$ & 0.263342 & 0.251385 & 0.250385 & 0.254097 & 0.250916 \\
		$t_{(00\frac{1}{2})}^{z z}$ & -0.413269 & -0.416136 & -0.417995 & -0.418873 & -0.417938 \\
		$t_{(00\frac{1}{2})}^{x x}$ &  0.000437 &  0.000001 &  0.000091 & -0.000827 & -0.000906 \\
		$t_{(100)}^{x x}$ & -0.442817 & -0.460940 & -0.477479 & -0.494620 & -0.511621 \\
		$t_{(100)}^{z z}$ & -0.122626 & -0.119119 & -0.115749 & -0.112470 & -0.109135 \\
		$t_{(100)}^{x z}$ &  0.212010 &  0.211878 &  0.211504 &  0.210866 &  0.209497 \\
		$t_{(10\frac{1}{2})}^{x x}$ & -0.001026 & -0.001219 & -0.001526 & -0.001485 & -0.001685 \\
		$t_{(10\frac{1}{2})}^{x z}$ & -0.033615 & -0.034944 & -0.036460 & -0.037885 & -0.039462 \\
		$t_{(10\frac{1}{2})}^{z z}$ &  0.039244 &  0.039238 &  0.039090 &  0.038898 & 0.038807 \\
		$t_{(110)}^{x x}$ &  0.074050 &  0.075795 &  0.076848 &  0.078037 &  0.078893 \\
		$t_{(110)}^{z z}$ & -0.021513 & -0.021161 & -0.020978 & -0.020753 & -0.020356 \\
		$t_{(11\frac{1}{2})}^{x x}$ & 0.003667 & 0.003907 & 0.004215 & 0.004825 & 0.005300 \\
		$t_{(11\frac{1}{2})}^{z z}$ & 0.001746 & 0.001445 & 0.001455 & 0.001144 & 0.000698 \\
		$t_{(200)}^{x x}$ & -0.051020 & -0.053800 & -0.056179 & -0.059085 & -0.062117 \\
		$t_{(200)}^{z z}$ & -0.009686 & -0.010302 & -0.010852 & -0.011420 & -0.012308 \\
		$t_{(200)}^{x z}$ &  0.016726 &  0.017444 &  0.018209 &  0.019199 &  0.020451 \\
		$t_{(20\frac{1}{2})}^{x x}$ & -0.003311 & -0.003349 & -0.003369 & -0.003545 & -0.003645 \\
		$t_{(20\frac{1}{2})}^{z z}$ & -0.004803 & -0.005019 & -0.005475 & -0.005933 & -0.00641 \\
		$t_{(20\frac{1}{2})}^{x z}$ &  0.003491 &  0.003744 &  0.003964 &  0.004285 & 0.004779 \\
		$t_{(210)}^{x x}$ & -0.004737 & -0.004792 & -0.004711 & -0.004814 & -0.004814 \\
		$t_{(210)}^{z z}$ & -0.000759 & -0.000680 & -0.000695 & -0.000538 & -0.000421 \\
		$t_{(210)}^{x z}$ &  0.002223 &  0.002105 &  0.001893 &  0.001545 &  0.001233 \\
		$t_{(300)}^{x x}$ & -0.012850 & -0.013165 & -0.013874 & -0.014273 & -0.014993 \\
		$t_{(300)}^{z z}$ & -0.002167 & -0.001944 & -0.001592 & -0.001414 & -0.001117 \\
		$t_{(300)}^{x z}$ &  0.006197 &  0.006203 &  0.006142 &  0.006256 &  0.006273 \\
		\hline
		\hline
		\label{T2}
	\end{tabular}
\end{table}

    \begin{table}[htbp]
	\caption{On-site energies $\varepsilon_a$ and hopping integrals $t_\delta^{ab}$ of the bilayer two-orbital tight-binding model corresponding to LNO thin film under different compressive strains. Here, $x$ and $z$ denote the $3d_{x^2-y^2}$ and $3d_{3z^2-r^2}$ orbital, respectively. Note that the vertical interlayer distance is assigned as $\frac12$. The unit of $\varepsilon_a$ and $t_\delta^{ab}$ is eV.}
	\centering
	\setlength{\tabcolsep}{15pt}
	\renewcommand{\arraystretch}{1.5}
	\begin{tabular}{cccccc}
		\hline
		\hline
		& $\epsilon$=1\%, $\epsilon_c$=0\% & $\epsilon$=2\%, $\epsilon_c$=0\% & $\epsilon$=3\%, $\epsilon_c$=0\% & $\epsilon$=4\%, $\epsilon_c$=0\% & $\epsilon$=5\%, $\epsilon_c$=0\% \\ \hline
		$\varepsilon_x$ & 0.592366 & 0.671492 & 0.749354 & 0.847526 & 0.954883 \\
		$\varepsilon_z$ & 0.271121 & 0.265276 & 0.263073 & 0.270183 & 0.279125 \\
		$t_{(00\frac{1}{2})}^{z z}$ & -0.418465 & -0.426960 & -0.435548 & -0.442643 & -0.450407 \\
		$t_{(00\frac{1}{2})}^{x x}$ &  0.000600 &  0.000114 & -0.000295 & -0.000490 & -0.000229 \\
		$t_{(100)}^{x x}$ & -0.442173 & -0.459573 & -0.475229 & -0.492042 & -0.508735 \\
		$t_{(100)}^{z z}$ & -0.125291 & -0.124333 & -0.123394 & -0.122272 & -0.121186 \\
		$t_{(100)}^{x z}$ &  0.213847 &  0.215794 &  0.217339 & 0.2184890 & 0.219043 \\
		$t_{(10\frac{1}{2})}^{x x}$ & -0.001056 & -0.001101 & -0.001211 & -0.001364 & -0.00157 \\
		$t_{(10\frac{1}{2})}^{x z}$ & -0.033859 & -0.035642 & -0.037368 & -0.039112 & -0.040918 \\
		$t_{(10\frac{1}{2})}^{z z}$ &  0.040132 &  0.040913 &  0.041842 &  0.042877 & 0.0438080 \\
		$t_{(110)}^{x x}$ &  0.073732 &  0.074952 &  0.075868 &  0.076741 &  0.077328 \\
		$t_{(110)}^{z z}$ & -0.021844 & -0.021975 & -0.022272 & -0.022267 & -0.022287 \\
		$t_{(11\frac{1}{2})}^{x x}$ &  0.003673 &  0.004131 & 0.004596 & 0.005083 & 0.005246 \\
		$t_{(11\frac{1}{2})}^{z z}$ &  0.001744 &  0.001753 & 0.001668 & 0.001246 & 0.000937 \\
		$t_{(200)}^{x x}$ & -0.050764 & -0.053063 & -0.055063 & -0.057675 & -0.060372 \\
		$t_{(200)}^{z z}$ & -0.009568 & -0.010011 & -0.010308 & -0.010915 & -0.011707 \\
		$t_{(200)}^{x z}$ &  0.016588 &  0.017084 & 0.0177390 &  0.018570 & 0.0195370 \\
		$t_{(20\frac{1}{2})}^{x x}$ & -0.003429 & -0.003549 & -0.003788 & -0.004057 & -0.004105 \\
		$t_{(20\frac{1}{2})}^{z z}$ & -0.004992 & -0.005415 & -0.005988 & -0.006433 & -0.006962 \\
		$t_{(20\frac{1}{2})}^{x z}$ &  0.003650 &  0.004085 &  0.004293 &  0.004671 &  0.005282 \\
		$t_{(210)}^{x x}$ & -0.004722 & -0.004643 & -0.004759 & -0.004680 & -0.004688 \\
		$t_{(210)}^{z z}$ & -0.000739 & -0.000664 & -0.000660 & -0.000586 & -0.000461 \\
		$t_{(210)}^{x z}$ &  0.002248 &  0.002053 &  0.001771 &  0.001522 &  0.001297 \\
		$t_{(300)}^{x x}$ & -0.012858 & -0.013182 & -0.013816 & -0.014257 & -0.015030 \\
		$t_{(300)}^{z z}$ & -0.002294 & -0.002117 & -0.001846 & -0.001634 & -0.001446 \\
		$t_{(300)}^{x z}$ &  0.006290 &  0.006248 &  0.006327 &  0.006385 &  0.006411 \\ \hline
		\hline
		\label{T1}
	\end{tabular}
\end{table}

\clearpage

\section{SINGULAR-MODE FUNCTIONAL RENORMALIZATION GROUP}
\subsection{SM-FRG flows of LNO film}

\begin{figure*}[htbp]
    \centering
    \includegraphics[width=0.85\linewidth]{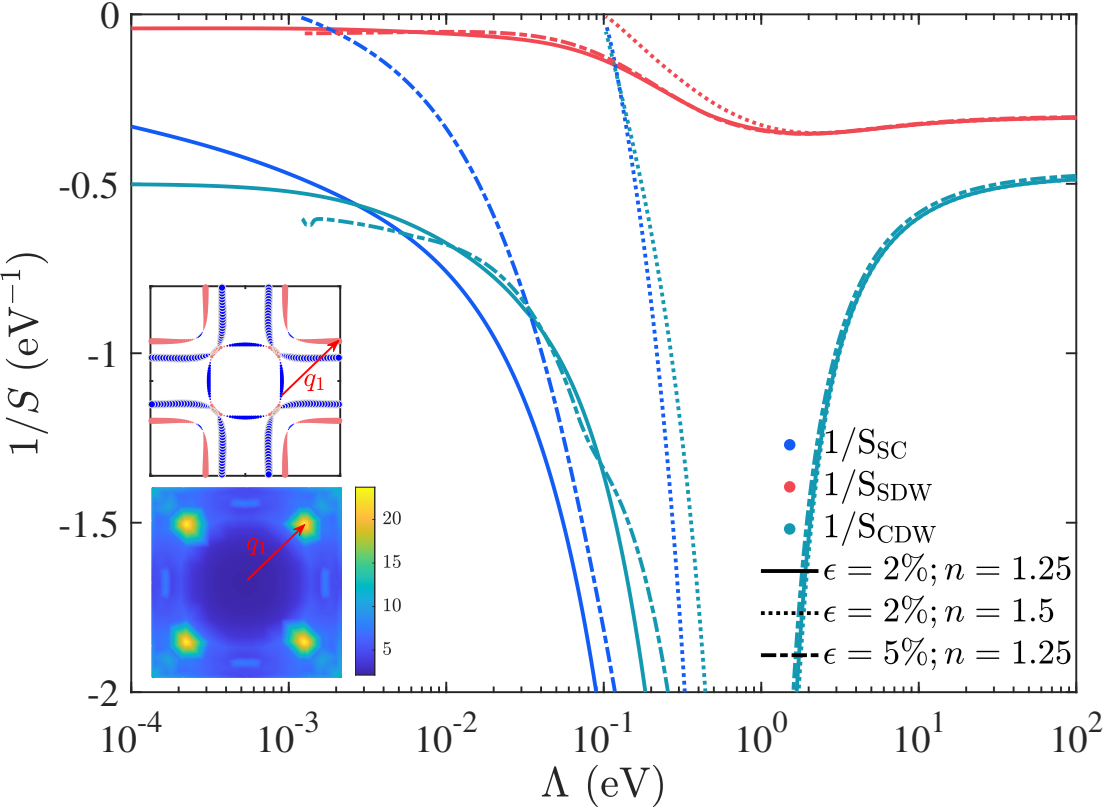}
    \caption{SM-FRG flows of $S^{-1}$ versus $\Lambda$ in the SC, SDW and CDW channels of LNO film, respectively, at $(\epsilon,n)$ = (2\%, 1.25), (5\%, 1.25) and (2\%, 1.5). The upper subset plots the gap function on the Fermi surfaces with the pink (blue) color indicating the positive (negative) sign and the size indicating the magnitude of the gap function. The lower subset presents the leading negative $|S(\0q)|$ in the SDW channel. Both the subsets are obtained at $(\epsilon,n)=(2\%, 1.25)$.}
    \label{flow}
\end{figure*}

\clearpage

\subsection{Real space superconducting pairing components}

The SC pairing function can be obtained directly from the fermion bilinear singular mode of the diverging SC channel. The SC pairings operator in real space is written as $H_{S C}=\sum_{i \delta, a b} \Delta_\delta^{a b} c_{i a \uparrow}^{\dagger} c_{i+\delta b \downarrow}^{\dagger}+h.c.$, where $a$ and $b$ denotes the orbital, $\delta = 0,1,2$ denote the on-site, on vertical bond and on in-plane nearest-neighbor bond, respectively. The primary superconducting pairing components are shown in Fig.~\ref{real_fig}, where $u_{0,1,2}$ and $v_{0,1,2}$ denotes $\Delta_\delta^{z z}$ and $\Delta_\delta^{x x}$, respectively. We also display the primary superconducting pairing components for ($U$,$J_H$) = (3,0.3) eV of LPNO and LNO films under different in-plane lattice constant compression ratios and $n=1.25$ in Table ~\ref{real_chen} and Table ~\ref{real_Hwang}, respectively.

\begin{figure*}[h]
	\includegraphics[width=0.3\textwidth]{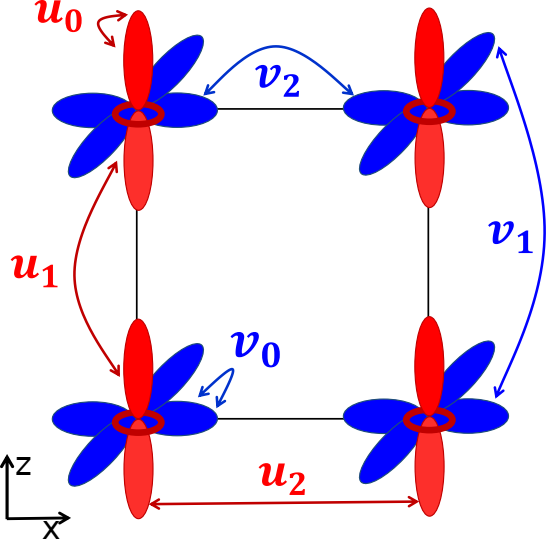}
	\caption{The real space superconducting pairing components up to translation and $C_{4v}$ symmetries.}
	\label{real_fig}
\end{figure*}

\begin{table}[htbp]
	\caption{Primary superconducting pairing components of LPNO film for ($U$,$J_H$) = (3,0.3) eV and $n=1.25$ under different in-plane lattice constant compression ratios.}
	\centering
	\setlength{\tabcolsep}{5pt}
	\renewcommand{\arraystretch}{1.5}
	\begin{tabular}{ccccccccc}
		\hline
		\hline
		& $u_0$ & $u_1$ & $u_2$ & $v_0$ & $v_1$ & $v_2$  \\ \hline

	$\epsilon$=2\%, $\epsilon_c$=1\% & -0.29303 & 0.37294 & -0.20121 & 0.25676 & 0.04816 & -0.02430  \\
	$\epsilon$=3\%, $\epsilon_c$=1.5\% & -0.33226 & 0.54933 & -0.12252 & 0.12190 & 0.03447 & -0.01098  \\
	$\epsilon$=4\%, $\epsilon_c$=2\% & -0.32034 & 0.58982 & -0.09699 & 0.07555 & 0.02647 & -0.00572  \\
	$\epsilon$=5\%, $\epsilon_c$=2.5\% & -0.28334 & 0.61586 & -0.09119 & 0.05027 & 0.02067 & -0.00317  \\
		\hline
		\hline
		\label{real_chen}
	\end{tabular}
\end{table}

\begin{table}[htbp]
	\caption{Primary superconducting pairing components of LNO film for ($U$,$J_H$) = (3,0.3) eV and $n=1.25$ under different in-plane lattice constant compression ratios.}
	\centering
	\setlength{\tabcolsep}{5pt}
	\renewcommand{\arraystretch}{1.5}
	\begin{tabular}{ccccccccc}
		\hline
		\hline
		& $u_0$ & $u_1$ & $u_2$ & $v_0$ & $v_1$ & $v_2$  \\ \hline

	$\epsilon$=3\%, $\epsilon_c$=0\% & -0.31397 & 0.37124 & -0.18666 & 0.27428 & 0.04246 & -0.02536  \\
	$\epsilon$=4\%, $\epsilon_c$=0\% & -0.34858 & 0.52732 & -0.12570 & 0.14490 & 0.03289 & -0.01283  \\
	$\epsilon$=5\%, $\epsilon_c$=0\% & -0.34505 & 0.57133 & -0.09825 & 0.09160 & 0.02659 & -0.00707  \\
		\hline
		\hline
		\label{real_Hwang}
	\end{tabular}
\end{table}

\end{widetext}

\bibliography{327film}

\end{document}